\documentclass[intlimits,twoside,a4paper]{article}
\usepackage{color}
\usepackage[cp1251]{inputenc}

\usepackage[eqsecnum]{cmpj3}

\newcommand{\be}{\begin{equation}}
\newcommand{\ee}{\end{equation}}
\newcommand{\bea}{\begin{eqnarray}}
\newcommand{\eea}{\end{eqnarray}}

\DeclareMathOperator{\Sp}{Sp}

\issue{2018}{21}{1}{13704}
\doinumber{10.5488/CMP.21.13704}

\title[Dynamic properties]{Dynamic properties of NH$_3$CH$_2$COOH$\cdot$H$_2$PO$_3$ ferroelectric}
\author[ I.R. Zachek, R.R. Levitskii, A.S. Vdovych,  O.B. Bilenka]{ I.R. Zachek\refaddr{label1}, R.R. Levitskii\refaddr{label2}, A.S. Vdovych\refaddr{label2},  O.B. Bilenka\refaddr{label1}
}
\addresses{\addr{label1} Lviv Polytechnic National
University, 12 Bandera St., 79013 Lviv, Ukraine
\addr{label2} Institute for Condensed Matter Physics of the
National Academy of Sciences of Ukraine,\\ 1 Svientsitskii St.,
79011 Lviv, Ukraine }

\date{Received November 15, 2017, in final form February 19, 2018}

\begin{document}

\maketitle

\begin{abstract}

Using a modified pseudospin model of NH$_3$CH$_2$COOH$\cdot$H$_2$PO$_3$ ferroelectric  taking into account the piezoelectric coupling with strains $\varepsilon_i$, $\varepsilon_4$, $\varepsilon_5$  and $\varepsilon_6$ within  Glauber method in two-particle claster approximation, we have calculated components of dynamic dielectric permittivity tensor and relaxation times of the model. At the proper set of theory parameters,  frequency and temperature dependences of the components of permittivity and temperature dependences of the relaxation times are studied.
 A satisfactory agreement of the theoretical results with experimental data for longitudinal permittivity is obtained.

\pacs 77.22.-d, 77.22.Ch, 77.22.Gm

\end{abstract}

\section{Introduction}

The problem of investigation of physical properties of ferroelectric materials has occupied one of the central places in condensed matter physics for a long time.  The presence of different classes of these materials with different crystal structure and chemical composition requires elaboration of universal methods for investigation of phase transition mechanisms. It is also necessary to develop  concrete microscopic theories for them, which could explain the  observed experimental data for thermodynamic and dynamic characteristics and anomalies in the behaviour of these characteristics in the phase transition region.

Granting this, glycinium phosphite NH$_3$CH$_2$COOH$\cdot$H$_2$PO$_3$ (GPI) is of special interest due to the combination of  structure elements typical of different classes of ferroelectric crystals.
In \cite{Stasyuk2003,Stasyuk2004,Stasyuk2004Ferro}  basing on the analysis of structural data  \cite{Taniguchi_JPSJ2003} it was determined that the main role in the phase transition in GPI is played by two structurally nonequivalent types of O-H$\ldots$O hydrogen bonds of different length, which connect phosphite groups  HPO$_3$ in the chains along the crystallographic $c$-axis. As a result, in  \cite{Stasyuk2003,Stasyuk2004Ferro} there was proposed a model of GPI crystal with proton ordering, within which the main peculiarities of its dielectric permittivity were explained qualitatively. Later, this model was supplement by taking into account the piezoelectric coupling of proton and lattice subsystems \cite{Zachek_PB2017}, which made it possible to calculate  thermal, piezoelectric and elastic characteristics of GPI.  At the proper set of theory parameters, a good agreement of the obtained theoretical results with corresponding experimental data for the crystals of this type was obtained.

In order to better understand the mechanism of phase transition in these crystals and  their physical properties, the effects of transverse electric fields \cite{Zachek_CMP2017} and uniaxial pressures \cite{Zachek_JPS2017} on the static physical properties of GPI  were calculated within the model proposed in  \cite{Zachek_PB2017}. A good agreement of the obtained theoretical results with the available experimental data was obtained.
This confirms the key role of proton ordering on the above mentioned bonds. It should be noted that several results obtained in these papers may be interpreted as predictions which will be a stimulus for further experimental investigations.

The aim of this paper is to study the relaxation phenomena in  GPI and explain the available experimental data  \cite{wie,tch,bar,Sobiestianskas1998} for  longitudinal dynamic characteristics within the proton ordering model proposed in  \cite{Zachek_PB2017}.

\section{Model of  GPI crystal}

\begin{figure}[!h]
\centering
\includegraphics[scale=0.5]{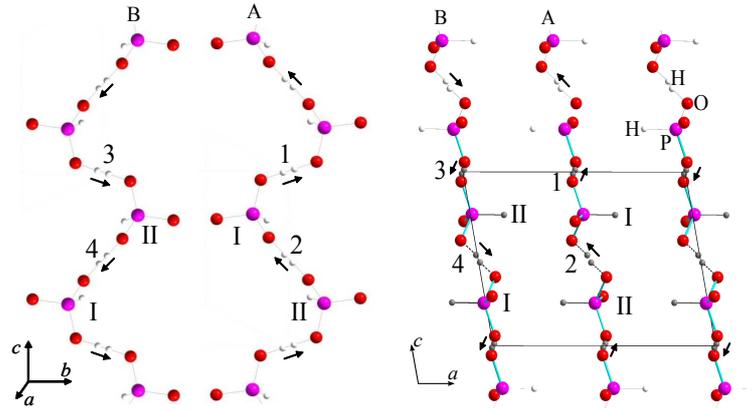}
\caption{(Colour online) Orientations of vectors ${\bf d}_{qf}$ in the primitive cell
in the ferroelectric phase \cite{Zachek_PB2017,Zachek_CMP2017}.} \label{struktura}
\end{figure}

The pseudospin model proposed in \cite{Zachek_PB2017} considers the system of protons in GPI, localized on O-H$\ldots$O bonds between phosphite groups HPO$_{3}$, which form chains along the crystallographic $c$-axis of the crystal (figure~\ref{struktura}).
Dipole moments ${\bf d} _{qf}={\pmb\muup}_f \frac{\sigma_{qf}}{2}$ are ascribed to the protons on the bonds. Here, \textit{q} is a primitive cell index, $f=1,\dots,4$; $\frac{\sigma_{qf}}{2}$ are pseudospin variables that describe the changes connected with reorientation of the dipole moments.

The Hamiltonian of a proton subsystem of GPI, which takes into account the short-range and long-range interactions and the applied electric fields $E_1$, $E_2$, $E_3$ along the positive directions of the Cartesian axes    $X$, $Y$ and $Z$ ($X \perp (b,c)$, $Y \parallel b$, $Z \parallel c$) can be written in such a way:
\bea
&& \hat H= N U_{\text{seed}} + \hat H_{\text{short}} + \hat H_{\text{long}} + \hat H_{\text E}\,, \label{H}
\eea
where $N$  is the total number of primitive cells.
The first term in (\ref{H}) is the ``seed'' energy, which relates  to the heavy ion sublattice and does not explicitly depend  on the configuration of the proton subsystem. It includes elastic, piezoelectric and dielectric parts expressed in terms of electric fields
$E_i$ and strains~$\varepsilon_i$:
\begin{align}
 U_{\text{seed}}&= v\Bigg[\frac{1}{2}\sum\limits_{i,i'=1}^3c_{ii'}^{E0}(T)\varepsilon_i \varepsilon_{i'}+
\frac{1}{2}\sum\limits_{j=4}^6c_{jj}^{E0}(T)\varepsilon_j^{2} +  \sum\limits_{i=1}^3c_{i5}^{E0}(T)\varepsilon_i\varepsilon_5 + c_{46}^{E0}(T)\varepsilon_4\varepsilon_6  \nonumber\\
 &-\sum\limits_{i=1}^3 e_{2i}^0 \varepsilon_i E_2 - e_{25}^0 \varepsilon_5 E_2 -
e_{14}^0 \varepsilon_4 E_1 - e_{16}^0 \varepsilon_6 E_1-  e_{34}^0 \varepsilon_4 E_3 - e_{36}^0 \varepsilon_6 E_3 
\nonumber\\
 &- \frac{1}{2}  \chi_{11}^{\varepsilon 0}E_1^2 - \frac{1}{2}
\chi_{22}^{\varepsilon 0}E_2^2 -
\frac{1}{2}  \chi_{33}^{\varepsilon 0}E_3^2- \chi_{31}^{\varepsilon 0}E_3E_1\Bigg]. 
\end{align}
Parameters $c_{ij}^{E0}(T)$, $e_{ij}^0$, $\chi_{ij}^{\varepsilon 0}$ are the so-called ``seed'' elastic constants,
piezoelectric stresses and dielectric susceptibilities, respectively; $v$ is the volume of a primitive cell.

The second term in  (\ref{H}) is the Hamiltonian of short-range interactions:
\bea
&& \hat H_{\text{short}} =  2w \sum\limits_{qq'} \left ( \frac{\sigma_{q1}}{2} \frac{\sigma_{q2}}{2} + \frac{\sigma_{q3}}{2}\frac{\sigma_{q4}}{2} \right)
\bigl( \delta_{{\bf R}_q{\bf R}_{q'}} + \delta_{{\bf R}_q + {\bf R}_{c},{\bf R}_{q'}} \bigr). \label{Hshort}
\eea
In (\ref{Hshort}),  $\sigma_{qf}$ is the $z$-component of the pseudospin operator that describes the state of the $f$-th bond ($f = 1, 2, 3, 4$) in the  $q$-th cell.
The first Kronecker delta corresponds to the interaction between protons in the chains near the tetrahedra HPO$_{3}$ of type ``I'' (figure~\ref{struktura}), where the second  one near the tetrahedra HPO$_{3}$ of type ``II'', ${\bf R}_{c}$ is the lattice vector along the crystallographic $c$-axis. Contributions into the energy of interactions between protons near the tetrahedra of different type, as well as the mean values of the pseudospins  $\eta_{f}=\langle\sigma_{qf}\rangle$, which are related to the tetrahedra of different type, are equal.
Parameter $w$, which describes the short-range interactions within the chains, is expanded linearly into the series over strains $\varepsilon_i$:
\be
w = w^{0} + \sum\limits_{i=1}^6 \delta_{i}\varepsilon_i. \label{w}
\ee

The third term in  (\ref{H}) describes  the long-range dipole-dipole interactions and indirect  (through the lattice vibrations)  interactions between protons, which are taken into account in the mean field approximation:
\bea
&&\hat H_{\text{long}} = \frac12 \sum_{qq'ff'}  J_{ff'}(qq') \frac{\langle \sigma_{qf}\rangle}{2}\frac{\langle \sigma_{q'f'}\rangle}{2}
- \sum_{qq'ff'}  J_{ff'}(qq') \frac{\langle \sigma_{q'f'}\rangle}{2}\frac{\sigma_{qf}}{2}.\label{Hlong}
\eea
Fourier transforms of interaction constants $J_{ff'} = \sum\nolimits_{q'} J_{ff'}(qq')$ at ${\bf k}=0$ are linearly expanded over the strains $\varepsilon_i$:
\begin{eqnarray}
&& J_{ff'} = J^0_{ff'} + \frac{\partial J_{ff'}}{\partial \varepsilon_i}\varepsilon_i
= J^0_{ff'} + \sum\limits_{i=1}^6\psi_{ff'i}\varepsilon_i.
\end{eqnarray}
Thus,  (\ref{Hlong}) can be written as:
\be \hat H_{\text{long}} = N H^{0}  - \sum\limits_q \sum\limits_{f=1}^4 {\cal H}_f \frac{\sigma_{qf}}{2}\,, \label{Hlongs}
\ee
where
\begin{align} 
\ H^{0}  &=
\frac18 J_{11}(\eta_1^2 + \eta_3^2) +\frac18 J_{22}(\eta_2^2  + \eta_4^2) + \frac14 J_{13}\eta_1\eta_3 + \frac14 J_{24}\eta_2\eta_4
  +\frac14 J_{12}(\eta_1\eta_2 + \eta_3\eta_4) \nonumber \\
  &   +\frac14 J_{14}(\eta_1\eta_4 + \eta_2\eta_3).
\label{Hs} 
 \end{align}
In  (\ref{Hlongs}) the following notations are used:
\bea && {\cal H}_1 = \frac{1}{2}J_{11}\eta_1 + \frac{1}{2}J_{12}\eta_2 +
\frac{1}{2}J_{13}\eta_3 + \frac{1}{2}J_{14}\eta_4\,, \nonumber\\
&& {\cal H}_2 = \frac{1}{2}J_{22}\eta_2 + \frac{1}{2}J_{12}\eta_1 +
\frac{1}{2}J_{24}\eta_4 + \frac{1}{2}J_{14}\eta_3\,,\nonumber \\
&& {\cal H}_3 = \frac{1}{2}J_{11}\eta_3 + \frac{1}{2}J_{12}\eta_4 +
\frac{1}{2}J_{13}\eta_1+ \frac{1}{2}J_{14}\eta_2\,, \nonumber\\
&& {\cal H}_4 = \frac{1}{2}J_{22}\eta_4 + \frac{1}{2}J_{12}\eta_3 +
\frac{1}{2}J_{24}\eta_2 + \frac{1}{2}J_{14}\eta_1. \label{H1234}
\eea

The fourth term in  (\ref{H}) describes the interactions of pseudospins with an external electric field:
\bea
&&\hat H_{\text E} = -\sum\limits_{qf} \pmb\muup_{f} {\bf E} \frac{\sigma_{qf}}{2}.\label{H_E}
\eea
Here, $\pmb\muup_{1}=(\mu_{13}^{x},\mu_{13}^{y},\mu_{13}^{z})$,  $\pmb\muup_{3}=(-\mu_{13}^{x},\mu_{13}^{y},-\mu_{13}^{z})$,  $\pmb\muup_{2}=(-\mu_{24}^{x},-\mu_{24}^{y},\mu_{24}^{z})$,  $\pmb\muup_{4}=(\mu_{24}^{x},-\mu_{24}^{y},-\mu_{24}^{z})$   are the effective dipole moments per one pseudospin.

The two-particle cluster approximation for short-range interactions is used for the calculation of thermodynamic characteristics of GPI. In this approximation, thermodynamic potential is given by:
\bea
G = N U_{\text{seed}} + NH^0     - k_{\text B} T  \sum\limits_q \bigg[ 2\ln \Sp \re^{-\beta \hat H^{(2)}_{q}}
- \sum\limits_{f=1}^4\ln \Sp \re^{-\beta \hat H^{(1)}_{qf}} \bigg]  - N v \sum\limits_{i=1}^6 \sigma_i \varepsilon_i. \label{G}
\eea
Here, $\beta=1/k_{\text B}T$, $k_{\text B}$ is Boltzmann constant, $\hat H^{(2)}_{q}$, $\hat H^{(1)}_{qf}$ are two-particle and one-particle Hamiltonians:
\bea
&& \hspace{-2ex} \hat H^{(2)}_{q} = - 2w \left( \frac{\sigma_{q1}}{2} \frac{\sigma_{q2}}{2} + \frac{\sigma_{q3}}{2}\frac{\sigma_{q4}}{2}\right) -  \sum\limits_{f=1}^4 \frac{y_f}{\beta}  \frac{\sigma_{qf}}{2}\,, \label{H2} \\
&& \hspace{-2ex}
\hat H^{(1)}_{qf} = - \frac{\bar y_f}{\beta}\frac{\sigma_{qf}}{2}\,, \label{H1}
\eea
where such notations are used:
\bea
&& \hspace{-2ex} y_f = \beta (  \Delta_f + {\cal H}_f + \pmb\muup_f {\bf E}),   \label{yf}\\
&& \hspace{-2ex}  \bar y_f =  \beta \Delta_f + y_f. \label{yfx}
\eea
The symbols $\Delta_f$ are the effective cluster fields created by the neighboring bonds
from outside the cluster. 
Minimizing the thermodynamic potential (\ref{G}) with respect to the cluster fields $\Delta_f$ and to the strains $\varepsilon_{i}$, and expressing  $\Delta_f$ through the equilibrium order parameters $\tilde{\eta}_{1}=\tilde{\eta}_{3}=\tilde{\eta}_{13}$, $\tilde{\eta}_{2}=\tilde{\eta}_{4}=\tilde{\eta}_{24}$, we have obtained a system of equations for the equilibrium order parameters and strains for the case of zero mechanical stresses and fields:
\begin{align}
\tilde{\eta}_{13}  &=  \frac{1}{\widetilde{D}}[\sinh (\tilde{y}_{13} + \tilde{\eta}_{24}) +
a^{2}\sinh (\tilde{y}_{13} - \tilde{\eta}_{24})
+2a\sinh \tilde{y}_{13}],   \nonumber \\
\tilde{\eta}_{24}  &=  \frac{1}{\widetilde{D}}[\sinh (\tilde{y}_{13} + \tilde{\eta}_{24}) -
a^{2}\sinh (\tilde{y}_{13}-\tilde{\eta}_{24})
+2a\sinh\tilde{\eta}_{24}], \nonumber \\
0  &=  c_{l1}^{E0}\varepsilon_1  +  c_{l2}^{E0}\varepsilon_2  +  c_{l3}^{E0}\varepsilon_3  +  c_{l5}^{E0}\varepsilon_5 -  \frac{2\delta_{l}}{v} +  \frac{2\delta_l}{v \tilde{D}}M_{\varepsilon}  \nonumber \\
&-  \frac{\psi_{1l}^+}{4v} \tilde{\eta}_{13}^{2} - \frac{\psi_{2l}^+}{2v} \tilde{\eta}_{13}\tilde{\eta}_{24}-\frac{\psi_{3l}^+}{4v} \tilde{\eta}_{24}^{2}\,, \qquad (l=1,\ldots,6),\label{sigma}
\end{align}
where such notations are used:
\begin{align}
 \tilde{y}_{13} &= \frac{1}{2} \ln \frac{1 +  \tilde{\eta}_{13}} {1 -  \tilde{\eta}_{13}}+ \beta\nu_{1}^{+} \tilde{\eta}_{13}+\beta\nu_{2}^{+}\tilde{\eta}_{24}\,,\qquad \tilde{y}_{24} = \beta\nu_{2}^{+} \tilde{\eta}_{13}+\frac{1}{2} \ln \frac{1 +  \tilde{\eta}_{24}} {1 -  \tilde{\eta}_{24}}+\beta\nu_{3}^{+}\tilde{\eta}_{24}\,, \nonumber\\
\nu_{l}^{\pm}&=\nu_{l}^{0\pm}+\sum\limits_{i=1}^6\psi_{li}^{\pm}\varepsilon_{i}\,, \qquad
 \nu_{1}^{0\pm}=\frac{1}{4}(J_{11}^{0}\pm J_{13}^{0}); \qquad \nu_{2}^{0\pm}=\frac{1}{4}(J_{12}^{0}\pm J_{14}^{0}); \qquad \nu_{3}^{0\pm}=\frac{1}{4}(J_{22}^{0}\pm J_{24}^{0});\nonumber\\
\psi_{1i}^{\pm}&=\frac{1}{4}(\psi_{11i}\pm \psi_{13i}), \qquad \psi_{2i}^{\pm}=\frac{1}{4}(\psi_{12i}\pm \psi_{14i}), \qquad \psi_{3i}^{\pm}=\frac{1}{4}(\psi_{22i}\pm \psi_{24i}),\nonumber \\
\tilde{D} &= \cosh (\tilde{y}_{13} + \tilde{y}_{24}) + a^{2}\cosh (\tilde{y}_{13} - \tilde{y}_{24}) + 2a\cosh \tilde{y}_{13} + 2a\cosh \tilde{y}_{24} +  a^{2} + 1, \nonumber\\
M_{\varepsilon}&=2a^{2}\cosh(\tilde{y}_{13}-\tilde{y}_{24})+ 2a \cosh \tilde{y}_{13}+2a\cosh \tilde{y}_{24}+2a^{2}.\nonumber
\end{align}

\section{Theoretical calculations of dynamic dielectric permittivity of me\-cha\-nically clamped  GPI crystal}

To calculate the dynamic properties we use an approach based on the ideas of a stochastic Glauber model
 \cite{gla}. Using the methods developed in  \cite{kn2009}, we obtain the  following system of Glauber equations for time dependent correlation functions of the pseudospins:
  \be
  - \alpha \frac{\rd}{\rd t} \Big\langle \prod_f \sigma_{qf} \Big\rangle   =
  \sum\limits_{f'}   \Big\langle \prod_f \sigma_{qf} \Big[1  -
  \sigma_{qf'} \tanh \frac12 \beta {\varepsilon}_{qf'}(t)\Big] \Big\rangle, \label{glauber}
  \ee
where parameter $\alpha$ determines the time scale of dynamic processes,  $\varepsilon_{qf'}(t)$ is the local field acting on the $f'$-th pseudospin
in $q$-th cell. We use a two-particle cluster approximation in order to obtain a closed system of equations. In this approximation, local fields $\varepsilon_{qf}(t)$ are coefficients at  $\sigma_{qf}/2$ in two-particle and one-particle Hamiltonians (\ref{H2}), (\ref{H1}). Correspondingly, these fields are presented in a two-particle approximation:
\bea
 \varepsilon_{q1}  =   w\sigma_{q2}  +  \frac{y_{1}}{\beta}\,, \quad \varepsilon_{q2}  =   w\sigma_{q1}  +  \frac{y_{2}}{\beta}\,,
  \quad \varepsilon_{q3}  =   w\sigma_{q4}  +  \frac{y_{3}}{\beta}\,, \quad \varepsilon_{q4}  =   w\sigma_{q3}  +  \frac{y_{4}}{\beta}\,,\label{eps_f}
\eea
and in a one-particle approximation:
 \bea
 \varepsilon_{qf}  =  \frac{\bar{y}_{f}}{\beta}.
  \eea

As a result, from (\ref{glauber}) we obtain a system of equations for mean values of pseudospins   $ \langle \sigma_{qf} \rangle =\eta_{f}$ in a two-particle approximation:
\bea
&& \alpha \frac{\rd}{\rd t} \eta_{1}  =  - \eta_{1} +  P_{1}\eta_{2}  +  L_{1}\,, \quad \alpha \frac{\rd}{\rd t} \eta_{3}  =  - \eta_{3} +  P_{3}\eta_{4}  +  L_{3}\,,\nonumber\\
&& \alpha \frac{\rd}{\rd t} \eta_{2}  =   P_{2}\eta_{1} -  \eta_{2}  +  L_{2}\,, \quad \alpha \frac{\rd}{\rd t} \eta_{4}  =   P_{4}\eta_{3} -  \eta_{4}  +  L_{4} \label{deta2}
\eea
and in a one-particle approximation:
\bea
 \alpha \frac{\rd}{\rd t} \eta_{f} = - \eta_{f} + \tanh \frac{\bar y_{f}}{2}\,,\label{deta1}
\eea
where the following notations are used:
\bea
&& P_{f}  =  \frac12 \left[ \tanh \left( \frac{\beta w}{2}  +  \frac{y_{f}}{2} \right)  -  \tanh
\left(- \frac{\beta w}{2}  +  \frac{y_{f}}{2} \right) \right],  \nonumber \\
&& L_{f}  =  \frac12 \left[ \tanh \left( \frac{\beta w}{2}  +  \frac{y_{f}}{2} \right)  +  \tanh
\left(- \frac{\beta w}{2}  +  \frac{y_{f}}{2} \right) \right].\nonumber
\eea

Let us restrict ourselves to the case of small deviations from equilibrium state to solve the equations~(\ref{deta2}) and (\ref{deta1}). For this case we write $\eta_{f}$ and effective fields $y_{f}$, $\bar y_{f}$ in the form of a sum of equilibrium values and their deviations from equilibrium values (a mechanically clamped crystal):
\begin{align}
\eta_{1,3} &= \tilde{\eta}_{13} + \eta_{1,3t}\,, \qquad \eta_{2,4} = \tilde{\eta}_{24} + \eta_{2,4t}\,, \nonumber\\
y_{1} &= \tilde{y}_{13} + y_{1t} =   \beta \big[\Delta_{13} + 2\nu_{1}^{+} \tilde{\eta}_{13} +2\nu_{2}^{+} \tilde{\eta}_{24}
+ \Delta_{1t} + \nu_{1}^{+}(\eta_{1t}+\eta_{3t}) + \nu_{2}^{+}(\eta_{2t}+\eta_{4t}) 
\nonumber\\
&+\nu_{1}^{-}(\eta_{1t}-\eta_{3t})+\nu_{2}^{-}(\eta_{2t}-\eta_{4t})
+\mu_{13}^{x}E_{1t} + \mu_{13}^{y}E_{2t} + \mu_{13}^{z}E_{3t}\big], \qquad E_{it} = E_i \re^{\ri\omega t},\nonumber \\
 y_{3} &= \tilde{y}_{13} + y_{3t} =   \beta \big[ \Delta_{13} + 2 \nu_{1}^{+} \tilde{\eta}_{13} +2 \nu_{2}^{+} \tilde{\eta}_{24}
+ \Delta_{1t} + \nu_{1}^{+}(\eta_{1t}+\eta_{3t}) + \nu_{2}^{+}(\eta_{2t}+\eta_{4t})\nonumber\\ 
&- \nu_{1}^{-}(\eta_{1t}-\eta_{3t})-\nu_{2}^{-}(\eta_{2t}-\eta_{4t})  
-\mu_{13}^{x}E_{1t} + \mu_{13}^{y}E_{2t} - \mu_{13}^{z}E_{3t} \big],\nonumber\\
y_{2} &= \tilde{y}_{24} + y_{2t} =   \beta\big[ \Delta_{24} + 2\beta \nu_{2}^{+} \tilde{\eta}_{13} +2\beta \nu_{3}^{+} \tilde{\eta}_{24}
+ \Delta_{2t} + \nu_{2}^{+}(\eta_{1t}+\eta_{3t}) + \nu_{3}^{+}(\eta_{2t}+\eta_{4t}) \nonumber\\
&+ \nu_{2}^{-}(\eta_{1t}-\eta_{3t})+\nu_{3}^{-}(\eta_{2t}-\eta_{4t}) 
-\mu_{24}^{x}E_{1t} - \mu_{24}^{y}E_{2t} +\mu_{24}^{z}E_{3t} \big],\nonumber\\
y_{4} &= \tilde{y}_{24} + y_{4t} =  \beta\big[ \Delta_{24} + 2\beta \nu_{2}^{+} \tilde{\eta}_{13} +2\beta \nu_{3}^{+} \tilde{\eta}_{24}
+ \Delta_{4t} + \nu_{2}^{+}(\eta_{1t}+\eta_{3t}) + \nu_{3}^{+}(\eta_{2t}+\eta_{4t})\nonumber\\
&- \nu_{2}^{-}(\eta_{1t}-\eta_{3t})-\nu_{3}^{-}(\eta_{2t}-\eta_{4t}) 
+\mu_{24}^{x}E_{1t} - \mu_{24}^{y}E_{2t} - \mu_{24}^{z}E_{3t}\big ],\nonumber\\
\bar y_f &=  \beta \Delta_f + \tilde y_f +  \beta \Delta_{ft}+ y_{ft}\,, \qquad \tilde y_1=\tilde y_3=\tilde y_{13}\,, \qquad \tilde y_2=\tilde y_4=\tilde y_{24}. \label{eta0t}
\end{align}
Here, $\Delta_{13}=\Delta_{1}=\Delta_{3}$, $\Delta_{24}=\Delta_{2}=\Delta_{4}$ are equilibrium effective cluster fields, and $\Delta_{ft}$ are their deviations from equilibrium values. Parameters $\nu_{i}^{\pm}$ describe long-range interactions.
We decompose the coefficients $P_{f}$ and $L_{f}$ in a series of $\frac{y_{ft}}{2}$ limited by linear items:

\bea
&&P_{1,3} = P_{13}^{(0)} + \frac{y_{1,3t}}{2} P_{13}^{(1)}, \qquad L_{1,3} = L_{13}^{(0)} + \frac{y_{1,3t}}{2} L_{13}^{(1)}, \nonumber\\
&&P_{2,4} = P_{24}^{(0)} + \frac{y_{2,4t}}{2} P_{24}^{(1)}, \qquad L_{2,4} = L_{24}^{(0)} + \frac{y_{2,4t}}{2} L_{24}^{(1)},\label{P0t}
\eea
where the following notations are used:
\begin{align}
P_{13}^{(0)} &= \frac{1-a^2}{Z_{13}}\,, \ \ P_{13}^{(1)}   =   -\frac{4a(1 - a^2)\sinh \tilde{y}_{13}}{Z_{13}^{2}}\,, \ \ L_{13}^{(0)} = \frac{2a \sinh \tilde{y}_{13}}{Z_{13}}\,,  \ \
L_{13}^{(1)}   =   \frac{4a[2a   +   (1  +  a^2) \cosh \tilde{y}_{13}]}{Z_{13}^{2}}\,, \nonumber\\
P_{24}^{(0)} &= \frac{1-a^2}{Z_{24}}\,,\ \  P_{24}^{(1)}   =    -\frac{4a(1 - a^2)\sinh \tilde{y}_{24}}{Z_{24}^{2}}\,, \ \ L_{24}^{(0)} = \frac{2a \sinh \tilde{y}_{24}}{Z_{24}}\,, \ \
L_{24}^{(1)}   =   \frac{4a[2a   +   (1  +  a^2)  \cosh \tilde{y}_{24}]}{Z_{24}^2}\,, \nonumber\\
Z_{13}&=1 + a^2  +  2a \cosh \tilde{y}_{13}; \ \ Z_{24}=1 + a^2  +  2a \cosh \tilde{y}_{24}\,,\nonumber\\
a &= \re^{-\frac{w}{k_{\text B}T}}, \ \ w = w^0 + \sum\limits_{i=1}^3\delta_i\varepsilon_i + \sum\limits_{j=4}^6\delta_j\varepsilon_j.\nonumber
\end{align}

Substituting (\ref{eta0t}), (\ref{P0t}) into (\ref{deta2}), (\ref{deta1}) and excluding parameter $\Delta_{ft}$, we obtained the following differential equations for sums and differences of proton unary distribution functions:
\bea
 && \hspace{-4ex} \frac{\rd}{\rd t} \left(    \begin{array}{c}
          ({ \eta}_{1t} - { \eta}_{3t})_{1} \\
          ({ \eta}_{2t} - { \eta}_{4t})_{1}
          \end{array}
             \right)   =   \left(   \begin{array}{ccc}
          m_{11}^{-} &     -m_{12}^{-} \\
          -m_{21}^{-} &      m_{22}^{-}
          \end{array}
            \right)
\left(    \begin{array}{c}
          ({ \eta}_{1t} - { \eta}_{3t})_{1} \\
          ({ \eta}_{2t} - { \eta}_{4t})_{1}
          \end{array}
             \right) -  \beta E_{1t}
 \left(\begin{array}{c}
     m_{1}\mu_{13}^{x} \\
     -m_{2}\mu_{24}^{x}
       \end{array}
      \right),\label{deta1m3x}\\
&& \hspace{-4ex} \frac{\rd}{\rd t} \left(    \begin{array}{c}
          ({ \eta}_{1t}+{ \eta}_{3t})_{2} \\
          ({ \eta}_{2t}+{ \eta}_{4t})_{2}
          \end{array}
             \right)   =   \left(   \begin{array}{ccc}
          m_{11}^{+} &      -m_{12}^{+} \\
          -m_{21}^{+} &      m_{22}^{+}
          \end{array}
            \right)
\left(    \begin{array}{c}
          ({ \eta}_{1t}+{ \eta}_{3t})_{2} \\
          ({ \eta}_{2t}+{ \eta}_{4t})_{2}
          \end{array}
             \right) -  \beta E_{2t}
 \left(\begin{array}{c}
     m_{1}\mu_{13}^{y} \vspace{0.2mm}\\
     -m_{2}\mu_{24}^{y}
       \end{array}
      \right),\label{deta1p3y}\\
 && \hspace{-4ex} \frac{\rd}{\rd t} \left(    \begin{array}{c}
          ({ \eta}_{1t}-{ \eta}_{3t})_{3} \\
          ({ \eta}_{2t}-{ \eta}_{4t})_{3}
          \end{array}
             \right)   =   \left(   \begin{array}{ccc}
          m_{11}^{-} &      -m_{12}^{-} \\
          -m_{21}^{-} &      m_{22}^{-}
          \end{array}
            \right)
\left(    \begin{array}{c}
          ({ \eta}_{1t}-{ \eta}_{3t})_{3} \\
          ({ \eta}_{2t}-{ \eta}_{4t})_{3}
          \end{array}
             \right) -  \beta E_{3t}
 \left(\begin{array}{c}
     m_{1}\mu_{13}^{z} \\
     m_{2}\mu_{24}^{z}
       \end{array}
      \right),\label{deta1m3z}
 \eea
where
\bea
&& m_{11}^{\pm} = \frac{1}{\alpha}   \bigl( 1 - \beta\nu_{1}^{\pm}r_{13}K_{13} \bigr), \qquad m_{12}^{\pm}=\frac{1}{\alpha} \left[(1+ K_{13}) P_{13}^{(0)} + \beta \nu_{2}^{\pm}r_{13}K_{13}\right],\nonumber\\
&& m_{21}^{\pm}= \frac{1}{\alpha} \left[(1+ K_{24}) P_{24}^{(0)} + \beta \nu_{2}^{\pm}r_{24}K_{24}\right], \qquad m_{22}^{\pm} = \frac{1}{\alpha}   \bigl( 1 - \beta\nu_{3}^{\pm}r_{24}K_{24} \bigr),\nonumber\\
&& m_{1} = \frac{1}{\alpha} K_{13}r_{13}\,, \qquad m_{2} = \frac{1}{\alpha} K_{24}r_{24}\,,\nonumber\\
&& K_{13}  =  \frac{P_{13}^{(1)} \tilde{\eta}_{13} + L_{13}^{(1)}}{2r_{13}  -  \big[P_{13}^{(1)} \tilde{\eta}_{13} + L_{13}^{(1)}\big]}\,, \qquad r_{13}  =   1  - \bigl( \tilde{\eta}_{13} \bigr)^2, \nonumber\\
&& K_{24}  =  \frac{P_{24}^{(1)} \tilde{\eta}_{24}  +  L_{24}^{(1)}}{2r_{24}  -  \big[P_{24}^{(1)} \tilde{\eta}_{24}  +  L_{24}^{(1)}\big]}\,,  \qquad r_{24}  =   1  -  \bigl( \tilde{\eta}_{24} \bigr)^2.\nonumber
\eea
Solving the equations  (\ref{deta1m3x})--(\ref{deta1m3z}), we obtained time-dependent unary distribution function of protons.
The components of dynamic susceptibility of GPI clamped crystal can be written as:
\bea
&&  \chi_{11}(\omega) = \chi_{11}^{ 0} + \lim\limits_{E_{1t}\to 0} \frac{1}{\upsilon}\left[ \mu_{13}^{x} \frac{\rd ({ \eta}_{1t}-{ \eta}_{3t})_{1}}{\rd E_{1t}}-\mu_{24}^{x} \frac{\rd ({ \eta}_{2t}-{ \eta}_{4t})_{1}}{\rd E_{1t}}\right],\nonumber\\
&&  \chi_{22}(\omega) = \chi_{22}^{0} + \lim\limits_{E_{2t}\to 0} \frac{1}{\upsilon}\left[ \mu_{13}^{y} \frac{\rd ({ \eta}_{1t}+{ \eta}_{3t})_{2}}{\rd E_{2t}}-\mu_{24}^{y} \frac{\rd ({ \eta}_{2t}+{ \eta}_{4t})_{2}}{\rd E_{2t}}\right], \nonumber\\
&&  \chi_{33}(\omega) = \chi_{33}^{0} + \lim\limits_{E_{3t}\to 0} \frac{1}{\upsilon}\left[ \mu_{13}^{z} \frac{\rd ({ \eta}_{1t}-{ \eta}_{3t})_{3}}{\rd E_{3t}}+\mu_{24}^{z} \frac{\rd ({ \eta}_{2t}-{ \eta}_{4t})_{3}}{\rd E_{3t}}\right].\nonumber
\eea
The obtained susceptibilities consist of the ``seed'' part and two relaxational modes:
 \bea
&&  \chi_{ii}(\omega) = \chi_{ii}^{ 0} +\sum\limits_{l=1}^2 \frac{\chi_{l}^{ i}}{1 + \ri\omega \tau_{l}^{i}}\,, \qquad i=1,2,3 \rightarrow x,y,z,  \label{Xii}
\eea
where
\begin{align}
\chi_{l}^{i} &= \frac{\beta}{2v}\frac{\tau_{1}^{i}\tau_{2}^{i}}{\tau_{2}^{i}-\tau_{1}^{i}}  \big\{ (-1)^{l-1}\big[(\mu_{13}^{i})^{2}m_{1}+(\mu_{24}^{i})^{2}m_{2}\big]\nonumber \\
 &+(-1)^{l}\tau_{l}^{i}\big[ (\mu_{13}^{i})^{2}m_{1}m_{22}^{\gamma} + (\mu_{24}^{i})^{2}m_{2}m_{11}^{\gamma} - \mu_{13}^{i}\mu_{24}^{i}(m_{1}m_{21}^{\gamma} + m_{2}m_{12}^{\gamma})\big]\big\},  \label{Xi}
\end{align}
$\tau_{1,2}^{i}$ are relaxation times of the following form:
\bea
&&(\tau_{1,2}^{i})^{-1} =  \frac{1}{2}\left[(m_{11}^{\gamma}+m_{22}^{\gamma}) \pm \sqrt{(m_{11}^{\gamma}+ m_{22}^{\gamma})^{2}-4(m_{11}^{\gamma}m_{22}^{\gamma}-m_{12}^{\alpha}m_{21}^{\gamma})} \right].\label{taui}
\eea
In (\ref{Xi}), (\ref{taui}) $\gamma=\text{``+''}$ for $i=y$ and $\gamma=\text{``--''}$ for $i=x, z$.

Components of dynamic dielectric permittivity of proton subsystem of GPI is as follows:
\bea
&& \varepsilon_{ii}(\omega) = 1+4\piup\chi_{ii}(\omega).  \label{epsii}
\eea

\section{Comparison of numerical calculations with the experimental data. Discussion of the obtained results}

To calculate the temperature dependence of
dielectric, elastic, piezoelectric and thermal characteristics
of GPI we need to set certain values of the following parameters:
\begin{itemize}
\item parameters of the short-range interactions $w^{0}$;

\item parameters of the long-range interactions $\nu_{f}^{0\pm}$ ($f=1,2,3$);

\item deformational potentials $\delta_{i}$,   $\psi_{fi}^{\pm}$ ($f=1,2,3$; $i=1,\ldots,6$);

\item effective dipole moments
$\mu_{13}^{a}$; $\mu_{24}^{a}$; $\mu_{13}^{b}$; $\mu_{24}^{b}$; $\mu_{13}^{c}$; $\mu_{24}^{c}$;

\item ``seed'' dielectric susceptibilities $\chi_{ii}^{  0}$;

\item ``seed'' coefficients of piezoelectric stress $e_{ij}^0$;

\item ``seed''  elastic constants $c_{ij}^{E0}$.
\end{itemize}

The values of the present theory parameters are determined while studying the static properties of  GPI \cite{Zachek_PB2017}.
The optimal values of  long-range interactions  $\nu_{f}^{0\pm}$ are as follows: $\tilde \nu_1^{0+}=\tilde \nu_2^{0+}=\tilde \nu_3^{0+}=2.643$~K, $\tilde \nu_1^{0-}=\tilde \nu_2^{0-}=\tilde \nu_3^{0-}=0.2$~K, where $\tilde \nu_{f}^{0\pm}=\nu_{f}^{0\pm}/k_{\text B}$.
The determined parameter $w^{0}$  of the GPI crystal is  $w^0/k_{\text B}=820$~K.
The optimal values of the deformational potentials $\delta_{i}$ are
$\tilde\delta_{1}=500$~K,  $\tilde\delta_{2}=600$~K, $\tilde\delta_{3}=500$~K, $\tilde\delta_{4}=150$~K, $\tilde\delta_{5}=100$~K,
$\tilde\delta_{6}=150$~K; $\tilde\delta_{i}$=${\delta_{i}}/{k_{\text B}}$.
The optimal values of the $\psi_{fi}^{\pm}$ are as follows:
$\tilde\psi_{f1}^{+} = 87.9$~K,  $\tilde\psi_{f2}^{+} = 237.0$~K,  $\tilde\psi_{f3}^{+} = 103.8$~K,
$\tilde\psi_{f4}^{+} = 149.1$~K,  $\tilde\psi_{f5}^{+} = 21.3$~K,  $\tilde\psi_{f6}^{+} = 143.8$~K,  $\tilde\psi_{fi}^{-}=0$~K,
where  $\tilde\psi_{fi}^{\pm} =\psi_{fi}^{\pm}/{k_{\text B}}$.
The effective dipole moments in the paraelectric phase are equal to $\mu_{13}^{x}=0.4\cdot 10^{-18}$~esu$\cdot$cm; $\mu_{13}^{y}=4.02\cdot 10^{-18}$~esu$\cdot$cm; $\mu_{13}^{z}=4.3\cdot 10^{-18}$~esu$\cdot$cm; $\mu_{24}^{x}=2.3\cdot 10^{-18}$~esu$\cdot$cm; $\mu_{24}^{y}=3.0\cdot 10^{-18}$~esu$\cdot$cm; $\mu_{24}^{z}=2.2\cdot 10^{-18}$~esu$\cdot$cm. In the ferroelectric phase, the $y$-component of the first dipole moment is  $\mu_{13\text{ferro}}^{y}=3.82\cdot 10^{-18}$~esu$\cdot$cm.

In \cite{tch} the transition temperature is $T_{\text c}=223.6$~K, and one should multiply the parameters $w^{0}$, $\nu_{f}^{0\pm}$, $\delta_{i}$, $\psi_{fi}^{\pm}$, $\mu_{13}^i$, $\mu_{24}^i$ by the coefficient 0.994.

The volume of a primitive cell of GPI is $\upsilon = 0.601\cdot 10^{-21}$~cm$^3$.

Parameter $\alpha$ is determined from the condition of an agreement of theoretically calculated and experimentally obtained frequency dependences of  $\varepsilon_{22}(\omega)$. We consider that parameter $\alpha$ slightly changes with temperature:
\[
\alpha = [1.6 - 0.011(\Delta T)]\cdot 10^{-14}~\text{s}, \qquad \Delta T = T - T_{\text c}.
\]

The ``seed'' coefficients of piezoelectric stress, dielectric susceptibilities and elastic constants are as follows:\\
$e_{ij}^0 = 0.0~\frac{\text{esu}}{\text{cm}^2}$;
\quad $\chi_{11}^{  0} = 0.1$, \quad $\chi_{22}^{  0}= 0.403$, \quad $\chi_{33}^{0} = 0.5$;\\
$c_{11}^{0E}   =   26.91\cdot10^{10}~\frac{\text{dyn}}{\text{cm}^2}$\,,\quad 
$c_{12}^{E0}   =   14.5 \cdot 10^{10}~\frac{\text{dyn}}{\text{cm}^2}$\,,\quad 
$c_{13}^{E0}   =   11.64 \cdot10^{10}~\frac{\text{dyn}}{\text{cm}^2}$\,,\quad 
$c_{15}^{E0} = 3.91  \cdot10^{10}$~$\frac{\text{dyn}}{\text{cm}^2}$\,,\\
$c_{22}^{E0} = [64.99 -  0.04(T-T_{\text c})] \cdot10^{10}$~$\frac{\text{dyn}}{\text{cm}^2}$\,,\quad 
$c_{23}^{E0} = 20.38\cdot10^{10}~\frac{\text{dyn}}{\text{cm}^2}$\,,\quad 
$c_{25}^{E0} = 5.64  \cdot10^{10}$~$\frac{\text{dyn}}{\text{cm}^2}$\,, \\
$c_{33}^{E0} = 24.41\cdot10^{10}~\frac{\text{dyn}}{\text{cm}^2}$\,,\quad 
$c_{35}^{E0} = -2.84  \cdot10^{10}$~$\frac{\text{dyn}}{\text{cm}^2}$\,,\quad 
$c_{55}^{E0} = 8.54 \cdot 10^{10}~\frac{\text{dyn}}{\text{cm}^2}$\,,\\
$c_{44}^{E0} = 15.31 \cdot10^{10}~\frac{\text{dyn}}{\text{cm}^2}$\,,\quad 
$c_{46}^{E0} = -1.1 \cdot 10^{10}$~$\frac{\text{dyn}}{\text{cm}^2}$\,,\quad 
$c_{66}^{E0} = 11.88 \cdot 10^{10}~\frac{\text{dyn}}{\text{cm}^2}$.

Other components $c_{ij}^{E0}\equiv0$.

From expression (\ref{Xii}) we can see that there are two contributions into the components of dielectric permittivity tensor of GPI. Numerical analysis shows that only one contribution to the permittivities is determinative ($\chi_{2}^i\gg
\chi_{1}^i$).

\begin{figure}[!h]
\centering
   \includegraphics[scale=0.85]{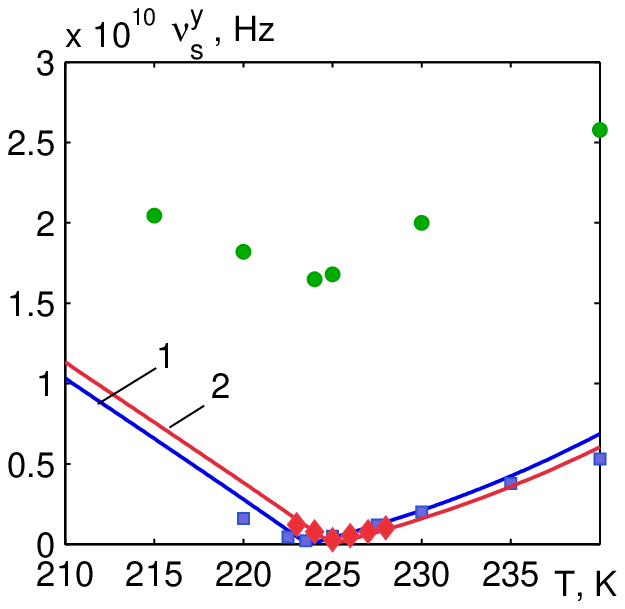}\qquad\includegraphics[scale=0.85]{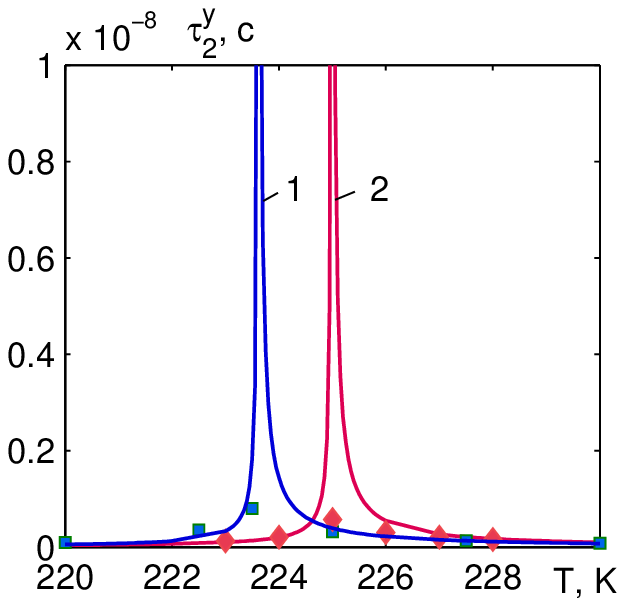}
\caption{(Colour online) The temperature dependence of relaxation frequency $\nu_{s}^{y}$: 1, ${\textcolor[rgb]{0.18,0.00,0.75}{\blacksquare}}$~---~\cite{tch}; 2, ${\textcolor[rgb]{1.00,0.00,0.00}{\blacklozenge}}$~---~\cite{ bar};  ${\textcolor[rgb]
{0.25,0.50,0.25}{\bullet}}$ --- \cite{Sobiestianskas1998}
 and relaxation time $\tau_{2}^{y}$: 1, ${\textcolor[rgb]{0.18,0.00,0.75}{\blacksquare}}$ ---
\cite{tch}; 2, ${\textcolor[rgb]{1.00,0.00,0.00}{\blacklozenge}}$ --- \cite{ bar}.} \label{tau}
\end{figure}

Let us first consider the longitudinal dynamic dielectric characteristics. They are predetermined by the behaviour of static  dielectric characteristics $\chi_{1}^y$, $\chi_{2}^y$ and relaxation times $\tau_{1}^{y}$, $\tau_{2}^{y}$  in the system. Relaxation time  $\tau_{2}^{y}$ is connected with some  relaxation frequency (soft relaxation mode) typical of this crystal
$\nu_{s}= (2\piup\tau_{2}^{y})^{-1}$, which conventionally separates the regions of low-frequency and high-frequency dispersion.
In figure~\ref{tau} there are presented temperature dependences of the relaxation frequencies  $\nu_{s}^{y}$, taken from
\cite{tch,bar,Sobiestianskas1998}, and the calculated temperature dependences of the longitudinal relaxation times $\tau_{2}^{y}=(2\piup\nu_{s}^{y})^{-1}$ \cite{tch,bar}.
Relaxation frequency, taken from \cite{Sobiestianskas1998}, greatly differs from the frequencies taken from  \cite{tch,bar}.
One can see from these figures that theoretical results satisfactorily agree with experimental data \cite{tch, bar}, except the phase transition region. Relaxation frequency $\nu_{s}^{y}$ decreases at approaching to the phase transition temperature and tend to zero at the temperature $T=T_{\text c}$. The calculated relaxation time $\tau_{2}^{y}$ has a singularity at  $T=T_{\text c}$, but experimental values of $\tau_{2}^{y}$ are finite at this temperature.

\begin{figure}[!t]
\centering
\includegraphics[scale=0.8]{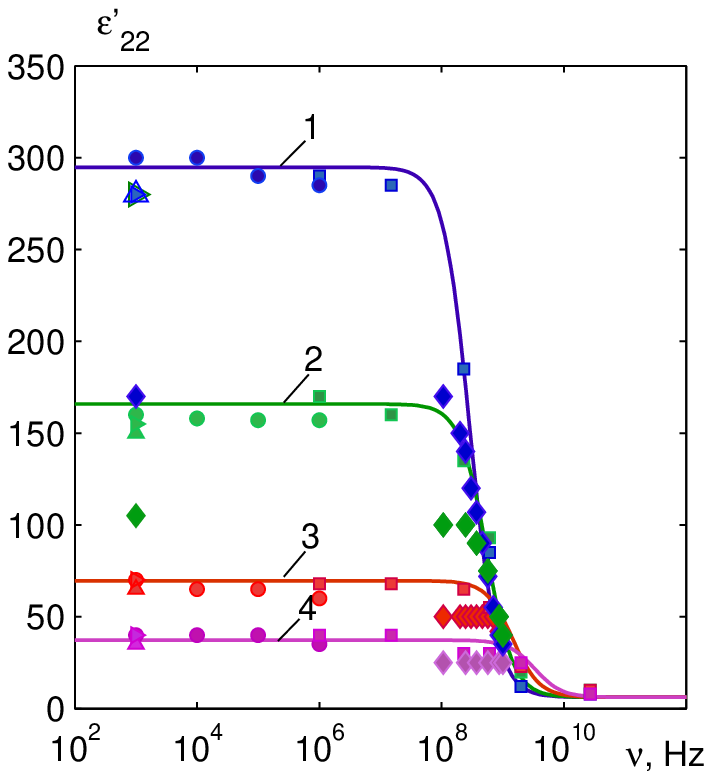}\qquad\includegraphics[scale=0.8]{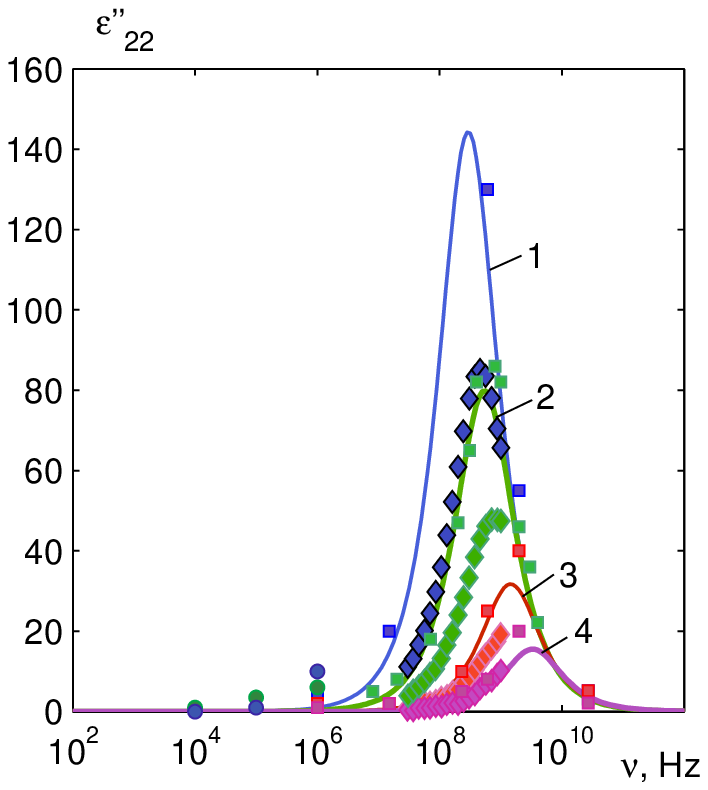}
\caption[]{(Colour online) The frequency dependences of real $\varepsilon'_{22}$ and imaginary $\varepsilon''_{22}$ parts of dielectric permittivity of  GPI
at different $\Delta T$(K): 1.0~---~1; 2.0~---~2; 5.0~---~3; 10.0~---~4; $\bullet$~\cite{wie}; $\blacksquare$~\cite{tch};
 $\blacklozenge$~\cite{bar}; $\blacktriangleright$~\cite{nay2}; $\blacktriangle$~\cite{dac}.} \label{eps22rf}
\end{figure}

\begin{figure}[!t]
\centering
 \includegraphics[scale=0.8]{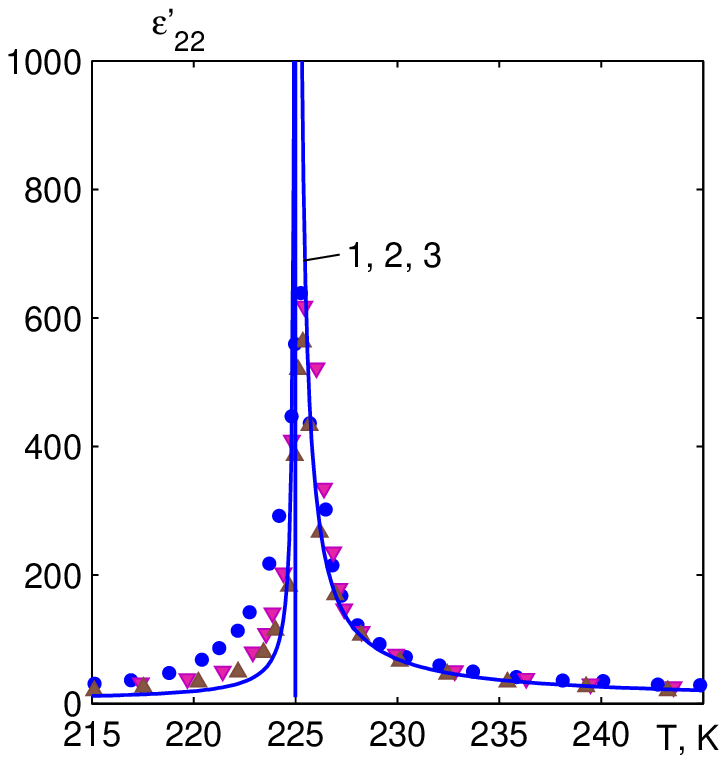}\qquad\includegraphics[scale=0.8]{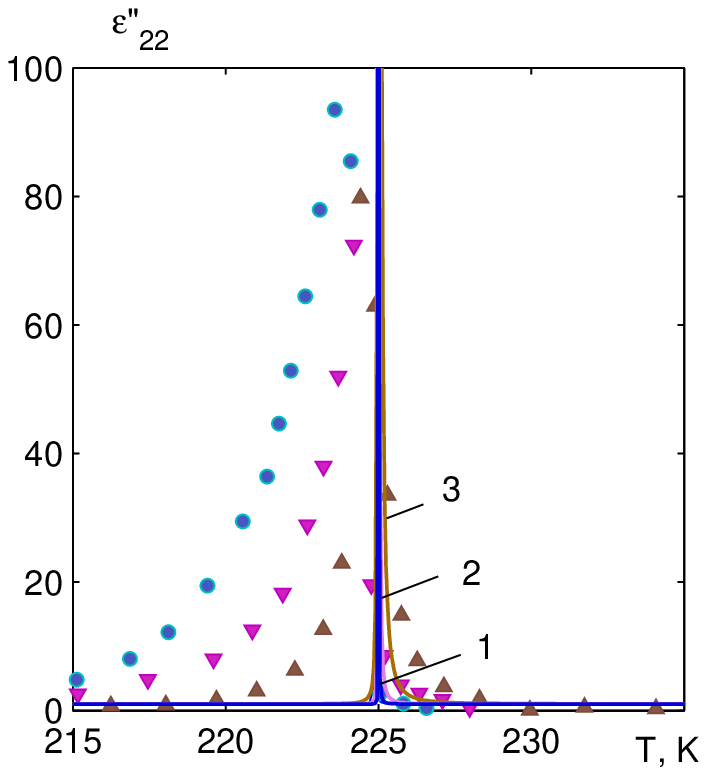} 
\caption[]{(Colour online) Temperature dependences of  $\varepsilon'_{22}$ and  $\varepsilon''_{22}$ of GPI
at different frequencies~$\nu$~(MHz): 0.01~---~1,
${\textcolor[rgb]{0.18,0.00,0.75}{\bullet}}$~\cite{wie}; 0.1~---~2, ${\textcolor[rgb]{0.75,0.20,0.75}{\blacktriangledown}}$~\cite{wie}; 1.0~---~3, ${\textcolor[rgb]{0.5,0.3,0.25}{\blacktriangle}}$ \cite{wie}.} \label{eps22rTW}
\end{figure}

At the frequencies  $\nu\ll\nu_{s}^{y}$ the real part of the dynamic dielectric permittivity  $\varepsilon_{22}'$ behaves as static, but the imaginary part  $\varepsilon_{22}'' $ is close to zero at all temperatures excepting the narrow region near $T_{\text c}$. One can see this on the frequency dependences  $\varepsilon_{22} (\nu)$  at different $\Delta T = T-T_{\text c}$ in the frequency region $\nu<10^7$~Hz (figure~\ref{eps22rf}), as well as  on the temperature dependences $\varepsilon_{22} (T)$  at low frequencies ($10^{4}$~Hz, $10^{5}$~Hz, $10^{6}$~Hz) (figure~\ref{eps22rTW}). 

\begin{figure}[!t]
\centering
 \includegraphics[scale=0.8]{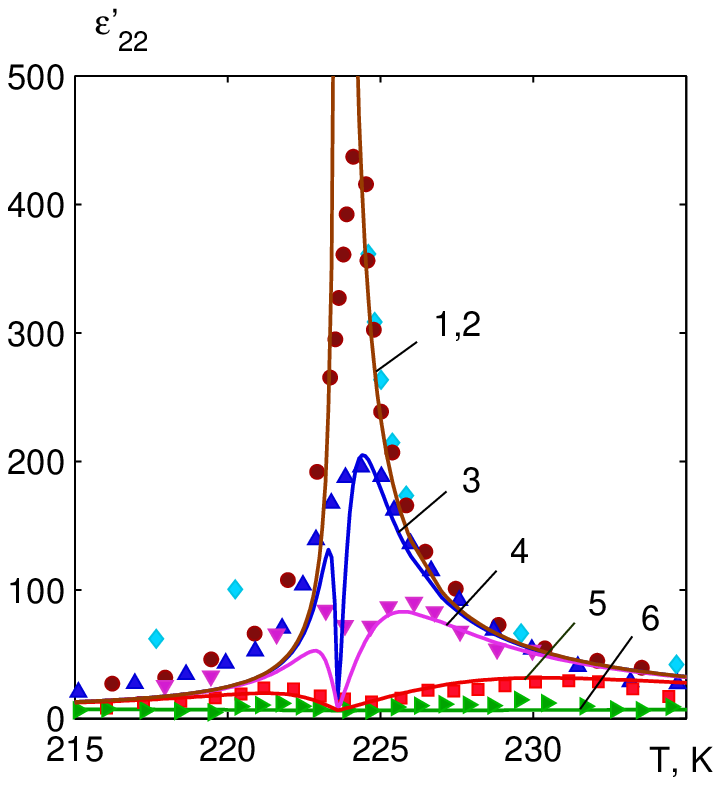}\qquad \includegraphics[scale=0.8]{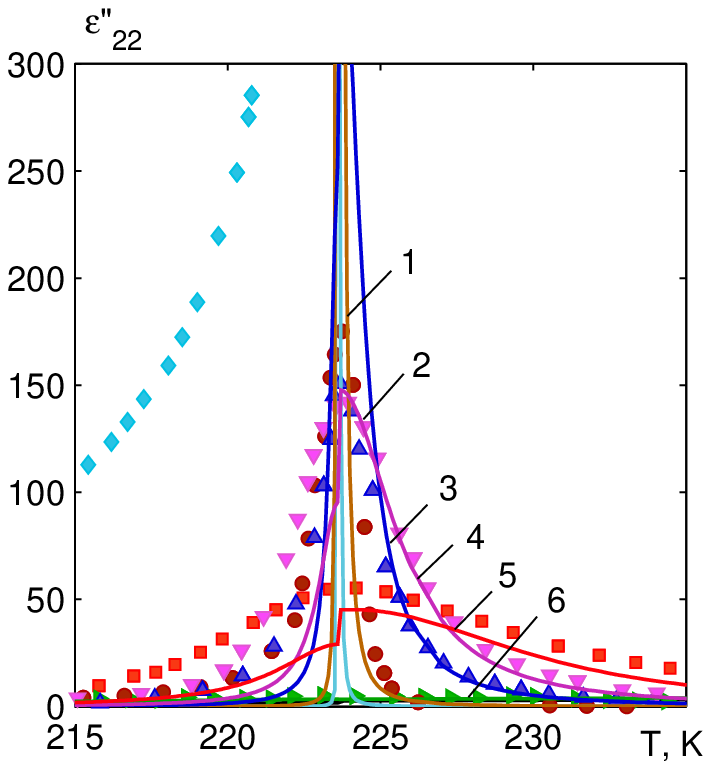}
\caption[]{(Colour online) Temperature dependences of  $\varepsilon'_{33}$ and $\varepsilon''_{33}$ of GPI crystal for various frequencies~$\nu$~(MHz):   1.0~---~1, ${\textcolor[rgb]{0.20,0.60,0.82}{\blacklozenge}}$~\cite{tch}; 15.0~---~2, ${\textcolor[rgb]
 {0.65,0.18,0.10}{\bullet}}$~\cite{tch}; 230.0~---~3, ${\textcolor[rgb]{0.18,0.00,0.75}{\blacktriangle}}$~\cite{tch}; 610~---~4, ${\textcolor[rgb]{0.75,0.20,0.75}{\blacktriangledown}}$~\cite{tch}; 2000~---~5, ${\textcolor[rgb]{1.00,0.00,0.00}{\blacksquare}}$~\cite{tch}; 27000~---~6, ${\textcolor[rgb]
{0.25,0.50,0.25}{\blacktriangleright}}$~\cite{tch}.} \label{eps22rTG}
\end{figure}

At the frequencies  $\nu \approx \nu_{s}$ we observe a relaxation dispersion, which reveals itself in the steep decreasing of the real part of dielectric permittivity  $\varepsilon_{22}' $ with an increasing frequency and in the large values of imaginary part $\varepsilon_{22}'' $;  the peak of $\varepsilon_{22}'' $ corresponds to the frequency $\nu_{s}$. One can see it on the frequency dependences $\varepsilon_{22} (\nu)$ at different $\Delta T = T-T_{\text c}$  in the frequency region  $10^7<\nu<10^{10}$~Hz (figure~\ref{eps22rf}), as well as  on the temperature dependences $\varepsilon_{22} (T)$  at the frequencies 1~MHz--27000~MHz (figure~\ref{eps22rTG}).

At the frequencies  $\nu \gg\nu_{s}$, the dielectric permittivity  behaves as a purely lattice contribution. It corresponds to the  frequency region  $\nu>10^{10}$~Hz on the frequency dependences $\varepsilon_{22} (\nu)$ in figure~\ref{eps22rf}.

An increase of the relaxation time $\tau_{2}^{y}$ and a decrease of the relaxation frequency  $\nu_{s}^{y}$ at approaching the temperature  $T=T_{\text c}$ manifests itself in the shift of the region of dispersion to lower frequencies in the frequency dependence  $\varepsilon_{22} (\nu)$ (figure~\ref{eps22rf}) at approaching the temperature  $T=T_{\text c}$, as well as in the availability of depression near $T=T_{\text c}$ on the temperature dependence $\varepsilon_{22}' (T)$, and of a sharp peak on the curve $\varepsilon_{22}'' (T)$ (figures~\ref{eps22rTW}, \ref{eps22rTG}). Since $\nu_{s}^{y} \rightarrow 0$ at $T=T_{\text c}$, then a  depression  of $\varepsilon_{22}' (T)$ and a peak of $\varepsilon_{22}'' (T)$ appears at all frequencies; they are very narrow at low frequencies and widen with an increase of frequency. The value of permittivity in the minimum point  (at $T=T_{\text c}$) is equal to the lattice contribution $\varepsilon_{22}^{  0}$. Since the experimental value is $\nu_{s}^{y} \neq 0$ at $T=T_{\text c}$, one can observe a low-frequency maximum in the experimental temperature dependence   $\varepsilon_{22}'(\nu,T)$ at low frequencies. Starting from frequency  $\nu_{s}\approx 10^{7}$, a depression-minimum appears instead of a maximum of $\varepsilon_{22}'(\nu,T)$, and this minimum decreases with an increase of frequency.

From figures~\ref{eps22rf}--\ref{eps22rTG} one can see that the proposed theoretical model satisfactorily describes the experimental data for the frequency and temperature dependences  $\varepsilon_{22}'(\nu,T)$ and $\varepsilon_{22}''(\nu,T)$ of GPI crystal in the paraelectric phase, with the exception of \cite{bar}, and less satisfactorily in the ferroelectric phase. A disagreement  of the theoretical curves with  the experimental data in the low-frequency region in the ferroelectric phase is connected with an essential role of domain processes in this region \cite{Czukwinski2001}, which are not taken into account in the proposed theory.

\begin{figure}[!t]
\centering
 \includegraphics[scale=0.8]{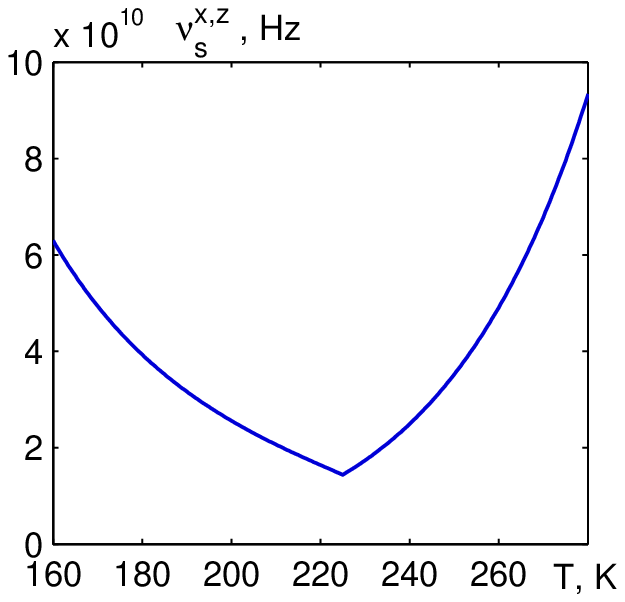} \qquad\includegraphics[scale=0.8]{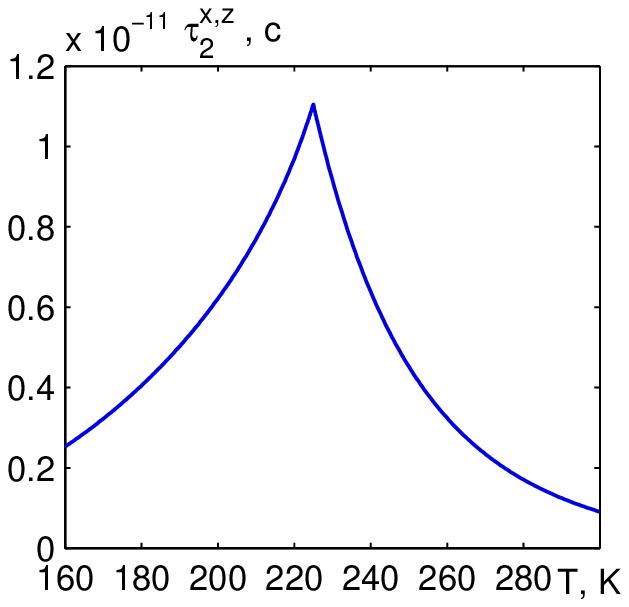}
\caption[]{(Colour online) The temperature dependences of relaxation frequencies $\nu_{s}^{x,z}$ and relaxation times $\tau_{2}^{x,z}$.} \label{tauxz}
\end{figure}

\begin{figure}[!t]
\centering
\includegraphics[scale=0.74]{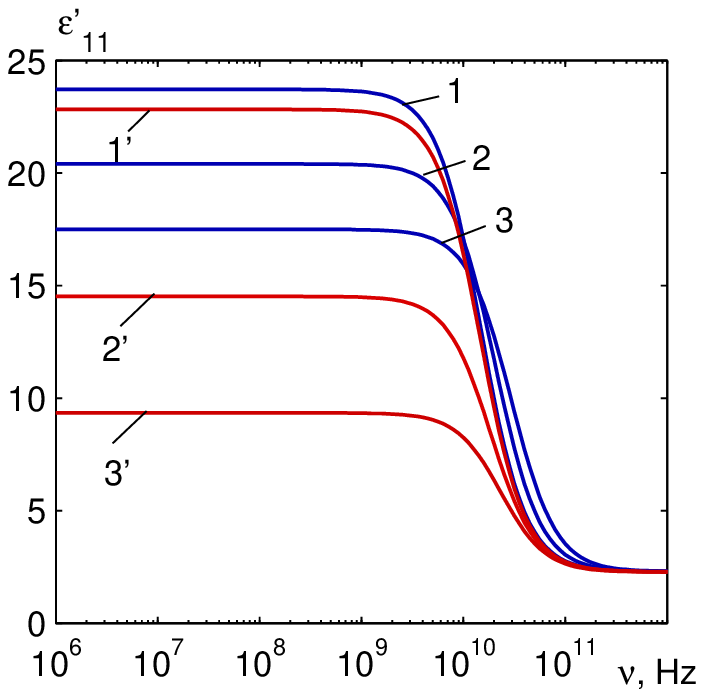}\qquad\includegraphics[scale=0.74]{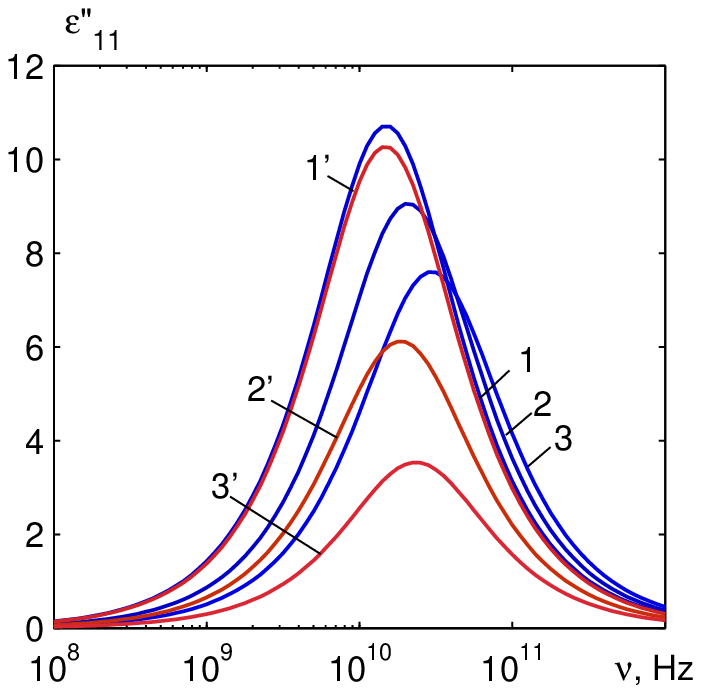}
\caption[]{(Colour online) Dispersion of real $\varepsilon'_{11}$ and imaginary $\varepsilon''_{11}$ parts of dielectric permittivity of GPI
 at different $\Delta T$~(K): 1~---~1; 10~---~2; 20~---~3; $-1$~---~$1'$; $-5$~---~$2'$; $-10$~---~$3'$.} \label{eps11rf}
\end{figure}

\begin{figure}[!t]
\vspace{-1mm}
\centering
\includegraphics[scale=0.74]{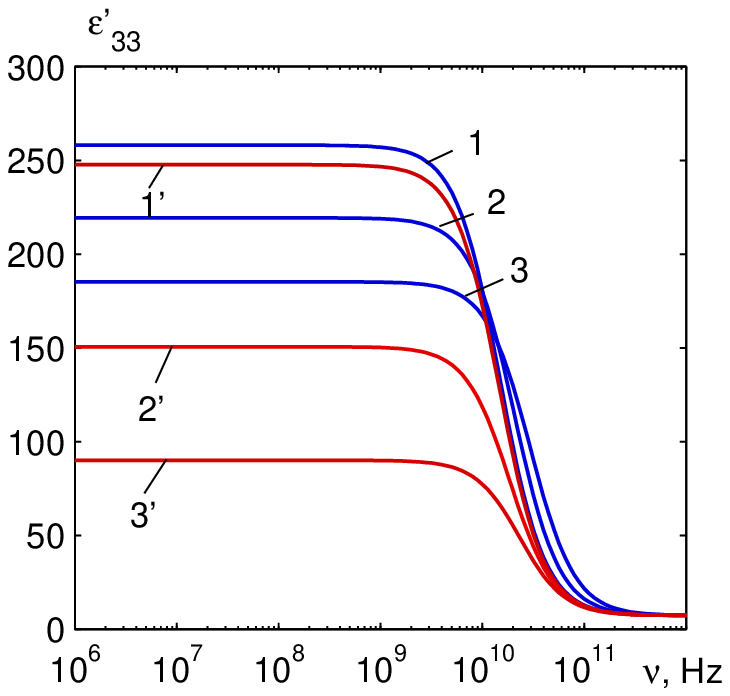}\qquad\includegraphics[scale=0.74]{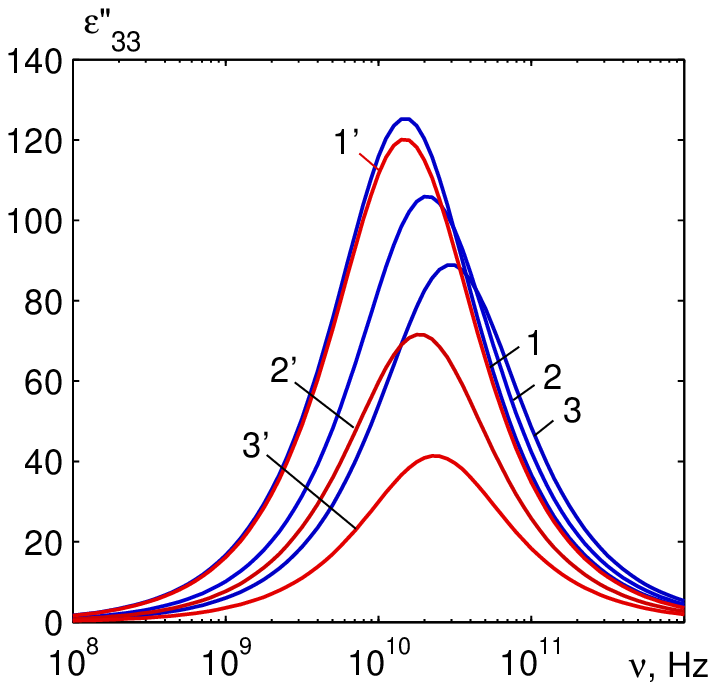}
\caption[]{(Colour online) Dispersion of real $\varepsilon'_{33}$ and imaginary $\varepsilon''_{33}$ parts of dielectric permittivity of GPI
 at different $\Delta T$~(K): 1~---~1; 10~---~2; 20~---~3; $-5$~---~$2'$; $-10$~---~$3'$.} \label{eps33rf}
\end{figure}

Let us discuss the transverse dynamic characteristics. Transverse relaxation frequencies  $\nu_{s}^{x,z}$ and
transverse relaxation  times $\tau_{2}^{x}$ and $\tau_{2}^{z}$ are calculated at the same  $\alpha$ as longitudinal  $\nu_{s}^{y}$ and
$\tau_{2}^{y}$. The frequencies  $\nu_{s}^{x,z}$ are higher than  $\nu_{s}^{y}$ and they also decrease at approaching the phase transition temperature (figure~\ref{tauxz}), and take on a nonzero value at $T=T_{\text c}$.
The transverse relaxation  times  $\tau_{2}^{x,z}$ in contrast to $\tau_{2}^{y}$ are finite at  $T=T_{\text c}$. This results in the frequency dependences of $\varepsilon_{11}(\nu)$ (figure~\ref{eps11rf}) and $\varepsilon_{33}(\nu)$ (figure~\ref{eps33rf}) at different $\Delta T$ that are qualitatively similar to the frequency dependences of  $\varepsilon_{22}(\nu)$, but the region of dispersion exists at higher frequencies and at weaker changes with temperature.

\begin{figure}[!t]
\vspace{-1mm}
\centering
 \includegraphics[scale=0.74]{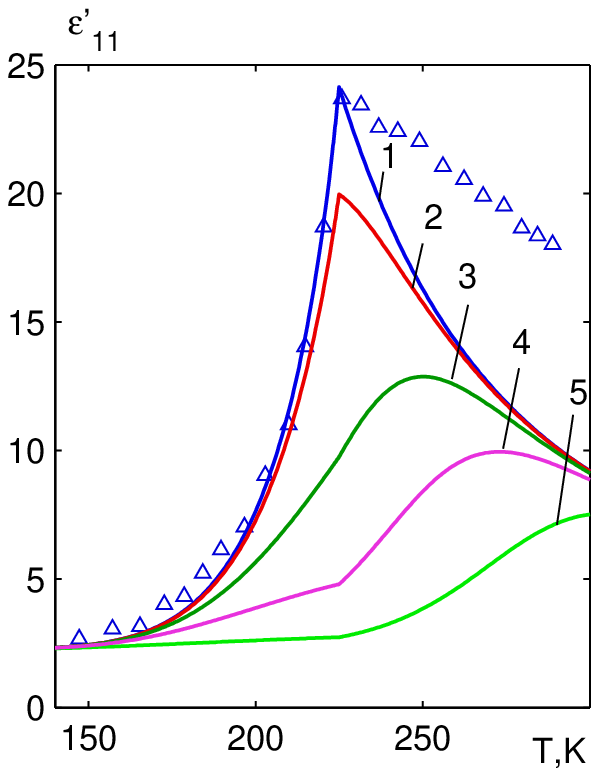} \qquad\includegraphics[scale=0.74]{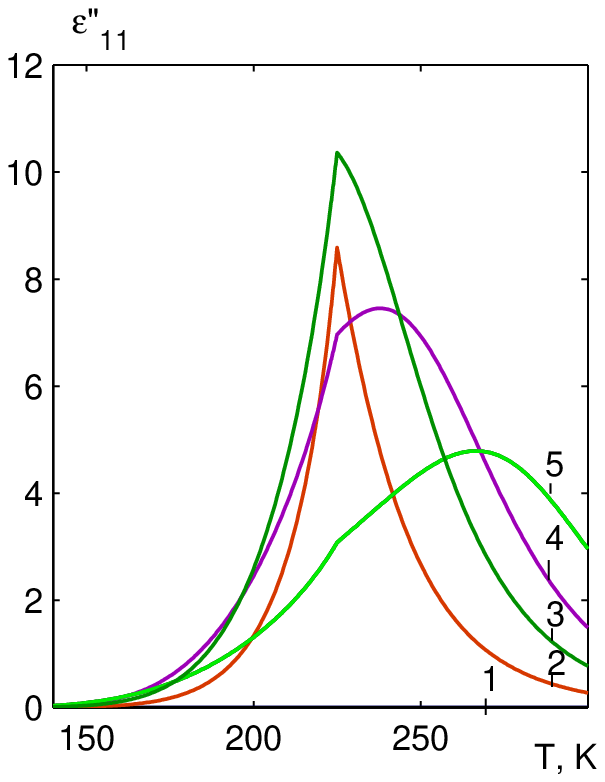}
\caption[]{(Colour online) Temperature dependences of  $\varepsilon'_{11}$ and $\varepsilon''_{11}$ for GPI crystal for various frequencies~$\nu$~(GHz):   0.0~---~1, ${\textcolor[rgb]{0.18,0.00,0.75}{\vartriangle}}$~\cite{dac}~(1~kHz); 7~---~2; 20~---~3; 40~---~4; 100~---~5.} \label{eps11rt}
\end{figure}

\begin{figure}[!t]
\centering
 \includegraphics[scale=0.74]{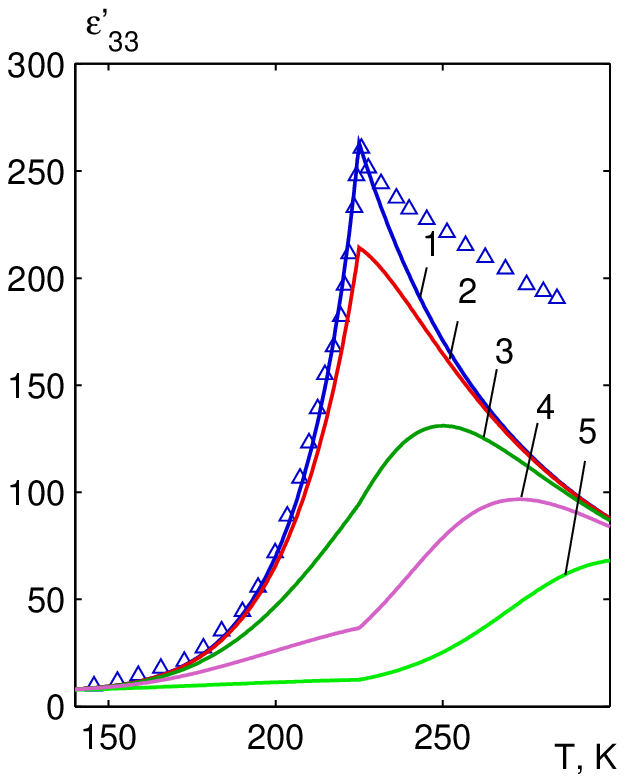} \qquad\includegraphics[scale=0.74]{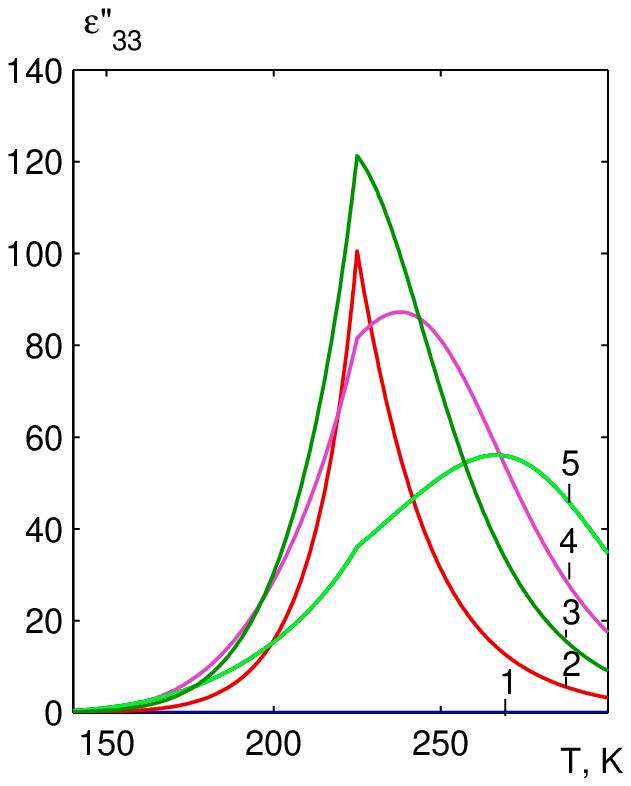}
\caption[]{(Colour online) Temperature dependences of  $\varepsilon'_{33}$ and $\varepsilon''_{33}$ for GPI crystal for various frequencies~$\nu$~(GHz):   0.0~---~1, ${\textcolor[rgb]{0.18,0.00,0.75}{\vartriangle}}$~\cite{dac}~(1~kHz); 7~---~2; 20~---~3; 40~---~4; 100~---~5.} \label{eps33rt}
\end{figure}

However, in the temperature dependences of $\varepsilon'_{11}$ and $\varepsilon'_{33}$, only the angle of the curve fracture in the point~$T_{\text c}$ changes (figures~\ref{eps11rt}, \ref{eps33rt}) instead of a depression near the phase transition temperature.
The maximum value of $\varepsilon'_{11,33}(T,\nu)$ at $T=T_{\text c}$ decreases with an increase of frequency.
Values of $\varepsilon''_{11,33}(T,\nu)$ at $T=T_{\text c}$ increase with an increase of frequency up to  $1.5\cdot10^{10}$~Hz. At higher frequencies, the maximum values of  $\varepsilon''_{11,33}(T,\nu)$ decrease and shift to the region of higher temperatures. Experimental investigations of transverse dynamic characteristics of GPI are very important to verify the obtained theoretical results in this regard. It is necessary to note that experimental data in figures~\ref{eps11rt} and \ref{eps33rt} are measured at frequency  1~kHz. They are close to static permittivities at such a small frequency.

\begin{figure}[!t]
\vspace{-1mm}
\centering
  \includegraphics[scale=0.78]{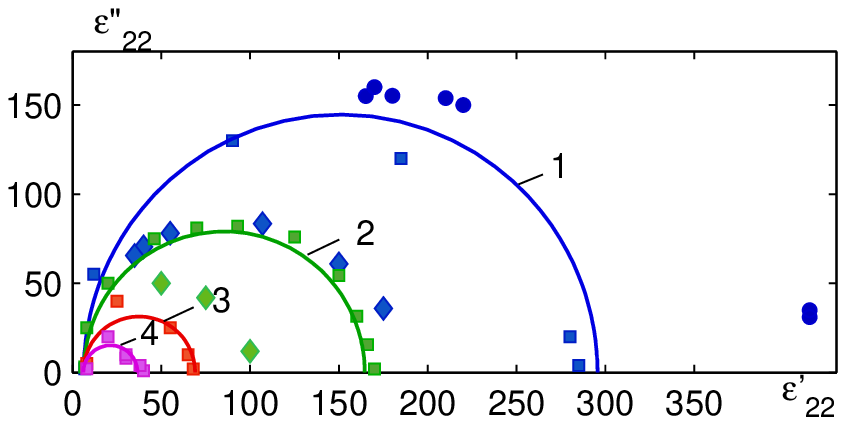}\\
 \includegraphics[scale=0.78]{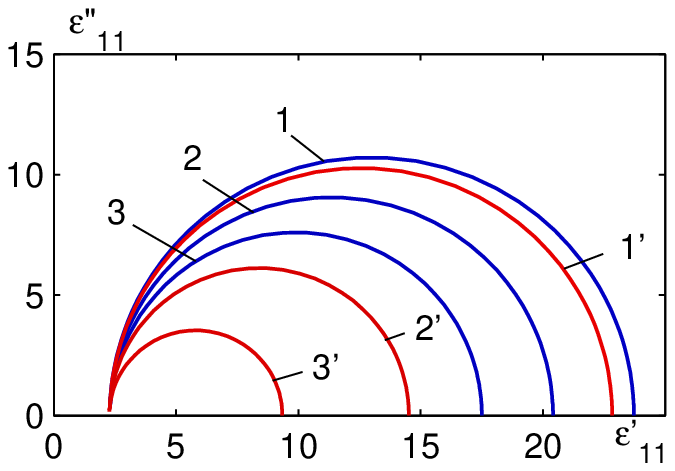} \qquad\includegraphics[scale=0.83]{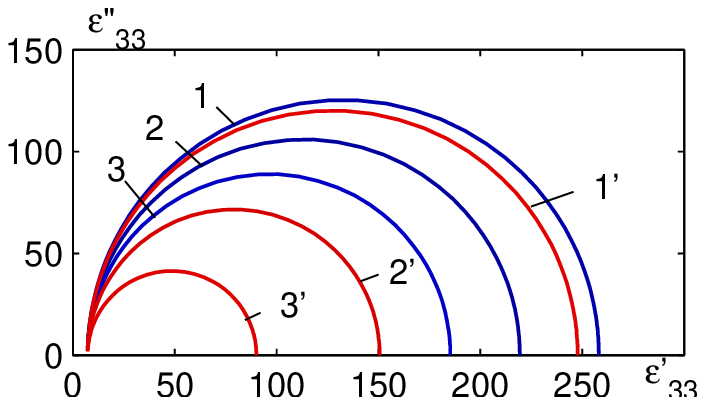}
\caption[]{(Colour online) Cole-Cole~(22) plot for GPI crystal at $\Delta T$~(K): 1~---~1,  ${\textcolor[rgb]{0.18,0.00,0.75}{\blacksquare}}$~\cite{tch}; ${\textcolor[rgb]{0.18,0.00,0.75}{\blacklozenge}}$\cite{bar}; ${\textcolor[rgb]{0.18,0.00,0.75}{\bullet}}$~\cite{Sobiestianskas1998}; 2~---~2,  ${\textcolor[rgb]{0.25,0.50,0.25}{\blacksquare}}$~\cite{tch}; ${\textcolor[rgb]{0.25,0.50,0.25}{\blacklozenge}}$~\cite{bar}; 5~---~3,  ${\textcolor[rgb]{1.00,0.00,0.00}{\blacksquare}}$~\cite{tch}; 10~---~4,  ${\textcolor[rgb]{0.75,0.20,0.75}{\blacksquare}}$~\cite{tch} and Cole-Cole~(11) and Cole-Cole~(33) plot at different $\Delta T$~(K): 1~---~1; 10~---~2; 20~---~3;  $-1$~---~$1'$; $-10$~---~$2'$; $-20$~---~$3'$.} \label{Coul}
\end{figure}

The results of calculation of Cole-Cole curves (figure~\ref{Coul}) witness for monodispersivity of dielectric permittivity in the crystals studied. The results of measurements of Cole-Cole curves for the longitudinal  permittivity, presented in \cite{tch,bar,Sobiestianskas1998}, disagree with each other. The calculated curves well agree with the results of  \cite{tch} for  longitudinal  permittivity.

\section{Conclusions}

Using the modified GPI model, the components of dynamic dielectric permittivity tensor and relaxation times are calculated in a two-particle claster approximation.
A satisfactory agreement of the theoretical results with experimental data for   longitudinal permittivity is obtained, with the exception of low-frequency region in the ordered phase,  inasmuch as the proposed theory does not take the domain processes into account, which can give a contribution into the above mentioned  frequency region.

It is determined that  the dynamic dielectric permittivity at low frequencies behaves as static; at the frequencies comparable with an inverse relaxation time, a relaxational dispersion is  observed; at high  frequencies, only a lattice contribution to permittivity reveals itself. The region of longitudinal dispersion in GPI shifts to the low frequencies at temperature approaching  the phase transition point, which is connected with a considerable increase of relaxation time at approaching the temperature $T_{\text c}$. The region of transverse dispersion lies at higher frequencies and weakly depends on temperature.

The obtained results for transverse characteristics bear the character of predictions and can  be a stimulus for further experimental investigations.

\ukrainianpart

\title{Динамічні властивості сегнетоелектрика NH$_3$CH$_2$COOH$\cdot$H$_2$PO$_3$}
\author{ І.Р. Зачек\refaddr{label1}, Р.Р. Левицький\refaddr{label2}, А.С. Вдович\refaddr{label2}, О.Б. Біленька\refaddr{label1}}
\addresses{\addr{label1} Національний університет ``Львівська політехніка'',  вул. С. Бандери, 12, 79013 Львів, Україна
\addr{label2} Інститут фізики конденсованих систем НАН України, вул. Свєнціцького, 1, 79011 Львів,  Україна
 }

\makeukrtitle

\begin{abstract}
\tolerance=3000%
Використовуючи модифіковану псевдоспінову модель сегнетоелектрика NH$_3$CH$_2$COOH$\cdot$H$_2$PO$_3$ шляхом врахування п'єзоелектричного зв'язку з деформаціями $\varepsilon_i$,
$\varepsilon_4$, $\varepsilon_5$, $\varepsilon_6$ в рамках методу Глаубера в
 наближенні двочастинкового кластера розраховано для неї компоненти тензора комплексної діелектричної проникності і часи
 релаксації. При належному виборі параметрів теорії вивчено частотні та температурні залежності компонент сприйнятливості
  та температурні залежності часів релаксації. Отримано задовільну згоду теоретичних результатів з експериментальними даними
  для поздовжньої проникності.
\keywords сегнетоелектрики, кластерне наближення, динамічна діелектрична проникність, час релаксації

\end{abstract}


\begin{thebibliography}{100}
	
\bibitem{Stasyuk2003} Stasyuk I., Czapla Z., Dacko S., Velychko O., Condens. Matter Phys., 2003, \textbf{6}, 483,
\bibdoi{10.5488/CMP.6.3.483}.

\bibitem{Stasyuk2004}  Stasyuk I., Czapla Z., Dacko S., Velychko O., J. Phys.: Condens. Matter, 2004, \textbf{16}, 1963,\\
\bibdoi{10.1088/0953-8984/16/12/006}.

\bibitem{Stasyuk2004Ferro}  Stasyuk I., Velychko  O., Ferroelectrics, 2004, \textbf{300}, 121,
\bibdoi{10.1080/00150190490443622}.

\bibitem{Taniguchi_JPSJ2003}  Taniguchi H., Machida  M., Koyano N., J. Phys. Soc. Jpn., 2003, \textbf{72}, 1111,
\bibdoi{10.1143/JPSJ.72.1111}.

\bibitem{Zachek_PB2017} Zachek I.R., Shchur Ya., Levitskii R.R., Vdovych A.S., Physica B, 2017, \textbf{520}, 164,\\
    \bibdoi{10.1016/j.physb.2017.06.013}.

\bibitem{Zachek_CMP2017} Zachek I.R., Levitskii R.R., Vdovych A.S., Stasyuk I.V., Condens. Matter Phys., 2017, \textbf{20}, 23706,
    \bibdoi{10.5488/CMP.20.23706}.

\bibitem{Zachek_JPS2017} Zachek I.R., Levitskii R.R., Vdovych A.S.,  J. Phys. Stud., 2017, \textbf{21}, 1704 (in Ukrainian).

\bibitem{wie} Wiesner M., Phys. Status Solidi B, 2003, \textbf{238}, 68,
    \bibdoi{10.1002/pssb.200301750}.

\bibitem{tch} Tchukvinskyi R., Czapla Z., Sobiestianskas R., Brilingas A., Grigas J., Baran J., Acta Phys. Pol. A, 1997, \textbf{92}, 1191,
    \bibdoi{10.12693/APhysPolA.92.1191}.

\bibitem{bar}  Baran J., Bator G., Jakubas R., Sledz M., J. Phys.: Condens. Matter, 1996, \textbf{8}, 10647,\\
    \bibdoi{10.1088/0953-8984/8/49/049}.
    
\bibitem{Sobiestianskas1998} Sobiestianskas R., Brilingas A., Czapla Z., J. Korean Phys. Soc., 1998, \textbf{32}, S377.

\bibitem{gla} Glauber R.J.,  J. Math. Phys., 1963, \textbf{4}, 294,
    \bibdoi{10.1063/1.1703954}.   

\bibitem{kn2009} Stasyuk I.V., Levitskii R.R., Moina A.P., Slivka A.G., Velychko O.V., Field and deformational effects in complex ferroelectric compounds, Grazhda, Uzhgorod, 2009, (in Ukrainian). 


\bibitem{nay2}  Nayeem J., Wakabayashi H.,  Kikuta T., Yamazaki T., Nakatani N., Ferroelectrics, 2002, \textbf{269}, 153,
    \bibdoi{10.1080/713716051}.


\bibitem{dac} Dacko S., Czapla Z., Baran J., Drozd M., Phys. Lett. A, 1996, \textbf{223}, 217,
    \bibdoi{10.1016/S0375-9601(96)00698-6}.


\bibitem{Czukwinski2001} Czukwinski R., Czapla Z., Styrkowiec R., Acta Phys. Pol. A, 2001, \textbf{100}, 897,
    \bibdoi{10.12693/APhysPolA.100.897}.
\end{thebibliography}
\end{document}